\documentclass[aps,prl,twocolumn,showpacs,superscriptaddress,groupedaddress]{revtex4}  
\usepackage{graphicx}  
\usepackage{dcolumn}   
\usepackage{bm}        
\usepackage{amssymb}   

\begin{document}

\widetext









\title{Analytical investigation of revival phenomena in periodically driven system }
\author{M. A. Shahzad\footnote{Email: shahzadma@gmail.com}}
\address{Department of Electronics, Quaid-i-Azam University Islamabad, Pakistan}

\begin{abstract}
We present theoretical study of revival phenomena for a wave packet initially well localized in a one-dimensional potential in the presence of an external periodic modulating field.  The classical motion, revival, and super-revival time scale are derived exactly for wave packet excited in the one dimension box potential. 
\end{abstract}
\maketitle

The phenomena of collapse and revivals [1,2] of wave packet has been investigated both theoretically and experimentally in a variety of physical system like that of Rydberg wave packets [3-5], wave packets in semiconductor quantum wells [6,7], coherent matter-wave in Bose-Einstein Condensation (BECs) [8,9],and  molecular systems [10,11]. These collapse and revivals occurred due to the time evolution of wave packet driven by discrete eigenvalue spectrum and depend on the quantum number $n$. For the physical system with a non-quadratic dependence on quantum number $n$ reveal a new sequence of collapse and revivals of the initially well localized wave packet. These are the super revivals [12] and is being observed in different physical systems [13]. \\
\indent In this article we present the time evolution and revivals of the wave packet in an arbitrary one-dimension potential in the presence of an external periodic field. The analytical analysis of this problem is explicitly discuss in [14,15]. In these articles the authors discuss the problem by truncating the eigenvalue quadratically. Keeping the energy term up to second order in the Taylor series expansion, they have mapped their initially unsolvable time-depended Schrodinger equation into known Mathieu equation whose solution are Floquet solution. However, the mention approached is valid for small modulation strength, i. e., $\lambda\ll 1$. Moreover, the Mathieu characteristic parameter are expanded up to second order while ignoring the reaming term under the approximation of small modulation strength. Here we consider the same problem by taking into account the energy term up to third order in the Taylor series expansion. Upon substituting into the Schrodinger equation, we obtained a third order differential equation. The solution of such equation is unknown. However, considering the term in the third derivative as a damping force, we again mapped our originally third order differential equation into known Mathieu-type equation. The procedure is valid and already discussed in [16]. \\
\indent We denote the unperturbed energy eigenstate and eigenvalues of the correspondent time-independent system by $|n\rangle$ and $E_n$, respectively, so that $H_0|n\rangle=E_n|n\rangle$, where $H_0$ is the unperturbed Hamiltonian. The Hamiltonian of the driven system in dimensionless form may be express as
\begin{equation}
H = H_0+\lambda V(x)\sin t
\end{equation}
where $\lambda$ is the modulation strength and $V(x)$ define the coupling.\\
\indent To obtained the solution for $|\Psi(t)\rangle$ to the time-dependent Schrodinger equation we used the ansatz [14], give by
\begin{equation}
\vert \psi(t)\rangle =\sum_n C_n(t)\vert n\rangle exp\Big\{-i\Big[E_r+(n-r){\hbar \over N}\Big]{t\over\hbar} \Big\}
\end{equation}
where $E_r$ is the average energy of the wave packet in the $N$th resonance.
Substituting the above equation into the Schrodinger wave equation and project the sate vector $\langle m \vert$ onto the result, we have
\begin{eqnarray}
i\hbar \dot{C}_m&=&\Big[E_m-E_r+(m-r){\hbar\over N} \Big]C_m(t)+{\lambda \over 2i}\sum_n V_{m,n}\nonumber \\
&&\times\Big\{e^{i(n-m+N)t/N}-e^{-i(n-m+N)t/N} \Big\}C_n(t)
\end{eqnarray}
where $V_{m,n}=\langle n\vert V(x)\vert m \rangle$ are the matrix elements of coupling $V(x)$. Under the approximation of keeping the stationary term $m=n\pm N$ and dropping the fast oscillating term, we obtained separate sets of coefficients. Moreover, assuming the matrix element independent of $n$, we can approximate by a constant $V$, i.e., $V_{m,m+N}\approx V_{m,m-N}=V$. Hence we get $N$ decoupled sets of equation:
\begin{equation}
i\hbar  \dot{C}_m=\Big[E_m-E_r+(m-r){\hbar\over N} \Big]C_m(t)+{\lambda \over 2i}\big(C_{m+N}-C_{m-N} \big)
\end{equation}
Expanding the energy up to $n$ terms , we have
\begin{eqnarray}
i\hbar  \dot{C}_m&=&(m-r)\big(E^{'}_r-{\hbar\over N} \big)C_m+\big({1\over 2 !}(m-r)^2 E^{''}_r+\cdot\cdot\cdot\cdot\nonumber \\
&&+{1\over n!}(m-r)^n E^{n(')_r} \big)C_m+{\lambda \over 2i}\big(C_{m+N}-C_{m-N} \big)\nonumber\\
\end{eqnarray}
we can write the above equation in angel representation by using the Fourier representation of $C_m$ as
\begin{eqnarray}
C_m&=&{1\over 2 \pi}\int^{2\pi}_{0}g(\phi)e^{-i(m-r)\phi}d\phi\nonumber\\
&=&{1\over 2N \pi}\int^{2N\pi}_{0}g(\theta)e^{-i(m-r)\theta/N}d\theta
\end{eqnarray}
with this particular choice of the Fourier representation, we obtained 
\begin{equation}
i\hbar \dot{g}(\theta,t)=\bar H g(\theta,t)
\end{equation}
where
\begin{eqnarray}
\bar H(\theta)&=&-{(i)^n N^n E^{n(')}_r\over n!}{\partial^n \over\partial \theta^n}-\cdot\cdot\cdot\cdot\nonumber\\
&& -{(i)^2 N^2 E^{''}_r\over 2!}{\partial^2 \over\partial \theta^2}-i(NE_r-\hbar){\partial\over\partial\theta}+\lambda V \sin(\theta)\nonumber\\
\end{eqnarray}
since the Hamiltonian $\bar H$ is time-independent, we can write the time evolution of $g(\theta,t)$ as
\begin{equation}
g(\theta,t)= \Theta(\theta)\exp({i\varepsilon t\over\hbar})
\end{equation}
using the above equation and considering the derivative up to third order, we have
\begin{eqnarray}
\Big({iN^3E^{'''}\over 6}{d^3\Theta(\theta)\over d\theta^3}+{N^3E^{''}\over 4}{d^2\Theta(\theta)\over d\theta^2}-i(NE^{'}-\hbar){d\Theta(\theta)\over d\theta}\nonumber\\
+\big(-\varepsilon+\lambda V \sin(\theta)\big)\Theta(\theta)\Big)=0\nonumber\\
\end{eqnarray}
Let
\begin{equation}
\alpha={iN^3E^{'''}\over 6} ,\qquad \beta= {N^3E^{''}\over 4}, \qquad \gamma= i(NE^{'}-\hbar)
\end{equation}
Eq. 10 can be rewritten as
\begin{equation}
\alpha{d^3\Theta(\theta)\over d\theta^3}+\beta{d^2\Theta(\theta)\over d\theta^2}-\gamma{d\Theta(\theta)\over d\theta}+\big(-\varepsilon+\lambda V \sin(\theta)\big)\Theta(\theta)=0
\end{equation}
\indent Let us suppose the term in third derivative is the damping force. Then we can assume the driven force to be harmonic of angular frequency $\omega$[16]. Hence we can used the approximation
\begin{equation}
{d^2\Theta\over d\theta}\approx-\big({i\gamma\over 2\beta}\big)^2\Theta
\end{equation}
Or
\begin{equation}
{d^2\Theta\over d\theta}\approx-\omega^2\Theta
\end{equation}
in the damping term. So that Eq. 12 becomes
\begin{equation}
\beta{d^2\Theta(\theta)\over d\theta^2}-\big(\alpha\omega^2+\gamma\big){d\Theta(\theta)\over d\theta}+\big(-\varepsilon+\lambda V \sin(\theta)\big)\Theta(\theta)=0
\end{equation}
\indent Using the transformation
\begin{eqnarray}
\Theta(\theta)=\Gamma(\theta)\exp\big({1\over 2\beta}(\alpha\omega^2+\gamma)\theta \big)
\end{eqnarray}
we have from Eq.(15)
\begin{eqnarray}
\ddot{\Gamma}(\theta)-{1\over 4 \beta^2}\Big\{(\alpha\omega^2+\gamma)^2+4\beta(\varepsilon+\lambda V\sin(\theta)) \Big\}\Gamma(\theta)&=&0\nonumber\\
\end{eqnarray}
by change of variable $\theta=2z+\pi/2$, we obtained a Mathieu equation
\begin{eqnarray}
\ddot{\Gamma}(z)+\Big\{-{1\over 4 \beta^2}(\alpha\omega^2+\gamma)^2-{\varepsilon\over\beta}+{\lambda V\over\beta}\cos(2z)\Big\}\Gamma(z)&=&0\nonumber\\
\end{eqnarray}
this is Mathieu-type equation with the quantity $-(1/4\beta^2)(\alpha\omega^2+\gamma)^2-(\varepsilon/\beta)$ acts as the Mathieu constant $a$ and the quantity $(\lambda V/\beta)$ acts as the Mathieu constant $q$.\\
\indent Hence under the approximation of considering the third derivative as a damping force we have mapped our equation into know Mathieu equation. The Floquet solution of the Mathieu equation [17] may be written as $\Gamma_\nu(z)=\exp(iz\nu)P_\nu(z)$, where $P_\nu(z)$ are the Mathieu function [18]. \\
\indent Thus, the Floquet solution $|\Psi_\kappa(t)\rangle$ to the Schrodinger equation can be written as
\begin{eqnarray}
|\psi_\kappa(t)\rangle=\exp(i\varepsilon_\kappa t/\kappa)|u_\kappa(t)\rangle
\end{eqnarray}
where $\varepsilon_\kappa$ and $|u_\kappa(t)\rangle$ are given by
\begin{eqnarray}
\varepsilon_\kappa=-{1\over 4\beta}(\alpha\omega^2+\gamma)^2-a_\kappa\beta
\end{eqnarray}
\begin{eqnarray}
|u
_\kappa(t)\rangle&=&{1\over 2\pi}\sum_{n}\exp(int/N)\nonumber\\
&&\times\int_{0}^{2\pi}d\phi e^{(i\nu N\phi/2)}e^{(-i(n-r)\phi/2)}P_{\nu(k)}|n\rangle\nonumber\\
\end{eqnarray}
\indent We calculate the autocorrelation function between the initial wave packet and the wave packet after certain evolution time $t$ to get the super revival time. Since the wave function given in Eq.(19) form a complete set of basis vector, therefore we can write
\begin{equation}
\langle\psi(0)|\Psi(t)\rangle=\sum_{n}|\xi_n|^2e^{i\varepsilon_n t/\hbar}
\end{equation}
where $\xi_n$ describe the probability amplitude in the $n$th state. Expanding the quasi-energy of the system around the resonant level $r$, the expression for the autocorrelation function becomes,
\begin{eqnarray}
\langle\psi(0)|\Psi(t)\rangle&=&\sum_{n}|\xi_n|^2\exp\{-i[\omega^{(0)}+(n-r)\omega^{(1)}\nonumber\\
&&+(n-r)^2\omega^{(2)}+\cdot\cdot\cdot]t \}
\end{eqnarray}
where $\omega^{(j)}$ denotes the $j$th derivative of $\varepsilon_n$ with respect to $n$ at $n=r$. From the above expression one can readily obtained the classical and quantum period for the driven system, i.e.,
\begin{eqnarray}
T_{cl}\equiv{2\pi\over\omega^{(1)}}={2\pi\hbar\over\varepsilon^{(1)}_n |_{n=r}},\qquad T_{\lambda}\equiv{2\pi\over\omega^{(2)}}={4\pi\hbar\over\varepsilon^{(2)}_n |_{n=r}}
\end{eqnarray} 
moreover, the super revival times appear in correspondence with the cubic terms of Taylor series expansion, that is
\begin{eqnarray}
T_{sr}\equiv{2\pi\over\omega^{(3)}}={12\pi\hbar\over\varepsilon^{(3)}_n |_{n=r}}\end{eqnarray} 
\indent Considering the modulation term to be small, i.e., $\lambda\ll 1$, we can approximate the Mathieu characteristic parameter $a_\nu$ up to second order term in $q$. Therefore, Eq.(20) than becomes
\begin{eqnarray}
\varepsilon_n=\beta\Big(\nu_n^2+{q^2\over2}{1\over \nu^2_n-1}\Big)+ O(q^4)
\end{eqnarray}
taking into account the non-resonant  situation, we have the explicitly relation for the classically revival time, that is
\begin{eqnarray}
T_{cl}&=&T_{0}\Big[2\nu_r+{q^2\over 2}{-2\nu_r\over(\nu^2-1)^2}\Big]
\end{eqnarray}
where $T_{01}=2\pi\hbar/\beta$. Similarly we can obtained the expression for quantum revival, 
\begin{eqnarray}
T_{\lambda}&=&2T_{0}\Big[2+{q^2\over 2}{2(3\nu_r^2-1)\over(\nu^2_r-1)^3}\Big]
\end{eqnarray}
\indent Further, taking into account the cubic term of Taylor series we get super revival, 
\begin{eqnarray}
T_{sr}&=&6T_{0}\Big[{q^2\over 2}{36\nu_r^2\over(\nu^2_r-1)^4}\Big]
\end{eqnarray}
with this approached we analytically obtained the revival time in the presence of the an external periodically driven system.\\
\indent In summery, we examine explicitly the revival feature in wave packet dynamics of a particle confined in a one-dimensional potential in the presence of an external periodic modulation modulated field. We have derived a rather complicated but accurate relation that reveals the super revival time as a function of the modulated strength. The calculation presented in this paper are more precise because of considering third power of quantum number dependence of the energy eigenvalue.

\end{document}